\documentclass{emulateapj}
\usepackage{amsmath}
\usepackage{verbatim}
\usepackage{graphicx}

\def\rE{\,R_{\rm E}}
\def\lcOneRescale{1.9 }
\def\lcTwoRescale{1.4 }
\def\xip{\xi_{\rm p}}
\def\xin{\xi_{\rm n}}
\def\vvo{{\bf v}_{\rm o}}
\def\vvs{{\bf v}_{\rm s}}
\def\vvl{{\bf v}_{\rm l}}
\def\vvc{{\bf v}_{\rm CMB,l}}
\def\bluered{The dotted blue (dashed red) curve shows the results from only the LC1 (LC2) analysis. }
\def\dos{D_{\rm OS}}
\def\dol{D_{\rm OL}}
\def\dls{D_{\rm LS}}
\def\tE{t_{\rm E}}
\def\ts{t_{\rm s}}
\def\cm{\,{\rm cm}}
\def\kms{\,{\rm km\,s^{-1}}}
\def\mmass{\left<M/M_\sun\right>}
\def\deg{^\circ}

\begin{document}
\title{The Transverse Peculiar Velocity of the Q2237+0305 Lens Galaxy and the Mean Mass of Its Stars}

\author{Shawn Poindexter\altaffilmark{1},
	Christopher S. Kochanek\altaffilmark{1}}
\altaffiltext{1}{
Department of Astronomy and Center for Cosmology and AstroParticle Physics,
The Ohio State University, 
140 W 18th Ave, Columbus, OH 43210, USA,
(sdp,ckochanek)@astronomy.ohio-state.edu}

\begin{abstract}

Using 11-years of OGLE V-band photometry of Q2237+0305, we measure the
transverse velocity of the lens galaxy and the mean mass of its stars.
We can do so because, for the first time, we fully include the random motions
of the stars in the lens galaxy in the analysis of the light curves.
In doing so, we are also able to correctly account for the Earth's parallax motion and the
rotation of the lens galaxy, further reducing systematic errors.
We measure a lower limit on the transverse speed of the lens galaxy, $v_{\rm t} > 338\kms$
(68\% confidence) and find a preferred direction to the East.
The mean stellar mass estimate including a well-defined velocity prior is
$0.12 \leq \left<M/M_\sun\right> \leq 1.94$ at $68\%$ confidence, with a median
of $0.52~M_\sun$.
We also show for the first time that analyzing subsets of a microlensing light curve,
in this case the first and second halves of the OGLE V-band light curve,
give mutually consistent physical results.

\end{abstract}


\section{Introduction} \label{sec:intro}

Quasar microlensing provides a unique tool for studying the properties
of cosmologically distant lens galaxies and the structure of quasars
\citep[see][]{Wambsganss06}.
Each of the multiple images of the quasar passes
through the gravitational potential of the stars along the line-of-sight
in the lens galaxy.
These stars microlens each of the ``macro'' images, so the total magnification
of each quasar image is strongly affected by the lensing effects of the stars
and the size of the quasar emission region.
Since the observer, lens galaxy, stars, and source quasar are all moving,
these magnifications change on timescales of 1-10 years with order unity amplitudes.

The relevant physical scale for quasar microlensing is the Einstein radius
projected into the source plane plane,
\begin{eqnarray}
\rE &=& D_{\rm OS} \sqrt{\frac{4G\left<M\right>}{c^2}\frac{D_{\rm LS}}{D_{\rm OL}D_{\rm OS}}} \nonumber \\
&=& 1.8\times 10^{17} \left(\frac{\left<M\right>}{M_\sun}\right)^{1/2}\cm,
\end{eqnarray}
where $G$ is the gravitational constant, $c$ is the speed of light,
$\left<M\right>$ is the mean stellar mass of the stars,
$D_{\rm LS}, D_{\rm OL}$, and $D_{\rm OS}$ are the angular diameter distances between
the lens-source, observer-lens, and observer-source respectively,
where we have used the lens and source redshifts for Q2237+0305
\citep[$z_{\rm l}=0.0394\,,z_{\rm s}=1.685$,][Q2237 hereafter]{Huchra85}.
The observed microlensing amplitude is controlled by the ratio between
the source size, $R_{\rm V} \approx 6\times10^{15}\cm$ (in V-band, see our
companion paper \citet{Poindexter09}, hereafter Paper II)
and $\rE$, in the sense that
smaller accretion disks produce larger variability amplitudes than larger disks.
If a source is much larger than $\rE$, there is little change in the magnification.

The timescale for variability is determined by the relative velocities of the
observer, the lens, its stars, and the source.  Generally, the lens motions dominate
\citep{Kayser89},
leading to two characteristic timescales. There is the timescale to cross an Einstein radius,
\begin{eqnarray}
\tE &\approx& \frac{\rE}{v_{\rm lens}} \frac{(1+z_{\rm l})\dol}{\dos} \nonumber \\
&\approx& 8 \left(\frac{v_{\rm lens}}{462\kms}\right)^{-1}
\left(\frac{\rE}{2\times10^{17}\cm}\right) {\rm yr}, \nonumber \\
\label{eqn:te}
\end{eqnarray}
and there is the timescale to cross the source,
\begin{eqnarray}
\ts &\approx& \frac{R_{\rm V}}{v_{\rm lens}} \frac{(1+z_{\rm l})\dol}{\dos} \nonumber \\
&\approx& 0.4 \left(\frac{v_{\rm lens}}{462\kms}\right)^{-1}
\left(\frac{R_{\rm V}}{6\times10^{15}\cm}\right) {\rm yr}, \nonumber \\
\label{eqn:ts}
\end{eqnarray}
where $v_{\rm lens}$ is the expected transverse speed of the lens
for Q2237.
These timescales are also affected by the direction of motion relative to the
shear \citep{Wambsganss90}.
Variation is guaranteed on timescale $\tE$ and can be observed on timescale
$\ts$ if the magnification pattern locally has structure on the scale of $R_{\rm V}$.

Quantitative studies using quasar microlensing have exploded in the last few years.
Recent efforts have studied the relationships between accretion disk size and black hole
mass \citep{Morgan10}, size and wavelength
\citep{Anguita08,Bate08,Eigenbrod08b,Poindexter08,Floyd09,Mosquera09},
sizes of non-thermal (X-ray) and thermal emission regions
\citep{Pooley07,Morgan08a,Chartas09,Dai09}, and the dark matter fraction of the lens
\citep{Dai09,Pooley09,Mediavilla09}.
All these studies have used static magnification patterns which ignore
the random stellar motions in the lens galaxy.
However, the stellar velocity dispersions of lens galaxies are comparable to the peculiar
velocities of galaxies, and ignoring them can lead to biased results.
For example, the average direction of motion
of the source is the same for all images, but this coherence is limited by the random
motions of the stars.  With fixed patterns one must either overestimate the coherence,
by using the same direction of motion for each image, or underestimate
it by using independent directions for each image \citep[see][]{Kochanek07}.
In either case, it would be dangerous to attempt measurements that depend on
this coherence, such as disk shape and orientation, or the transverse peculiar
velocity of the lens.

\citet{Kundic93}, \citet{Schramm93}, and \citet{Wambsganss95}
considered the effects of random stellar
motions and found that these motions can also lead to shorter microlensing
time scales because the pattern velocities of the microlensing caustics
can be much higher than any physical velocities.
As a result, measurements based on static magnification patterns may underestimate
source sizes or mean masses or overestimate the transverse velocity in order to match the
effects created by the random stellar motions.
\citet{Wyithe00} showed that it is possible to statistically correct for
these effects and that the velocity correction can be up to 40\%
depending on the optical depth and shear.
Another benefit to dynamic magnification patterns is the ability to properly account for
the velocity of Earth around the Sun and the rotation of the lens galaxy
\citep{Tuntsov04}.

Dynamic magnification patterns are also important because they impart
a well-measured physical scale to the patterns.
All direct microlensing observations are in ``Einstein units'' where the
length scale is $\left<M\right>^{1/2}\cm$.
Determining masses, velocities, or source sizes requires some sort of
dimensional prior.  In our studies we have generally followed \citet{Kochanek04a}
and used velocity priors designed to mimic the combined effects of random
and ordered motion.  The reason for focusing on the velocity is that we know
two of the contributions, our velocity and the stellar velocities, and the
remaining peculiar velocities of the lens and source are truly random
variables for which we have reasonable priors from cosmological models.
Source sizes turn out to be little affected by the choice of priors
\citep[see the discussion in][]{Kochanek04a}, but estimates of the true
velocity and the mean stellar masses are affected.  Hopefully by including the true random
stellar motions we can further reduce the sensitivity of microlensing
results to such priors.

\begin{figure*}
\plotone{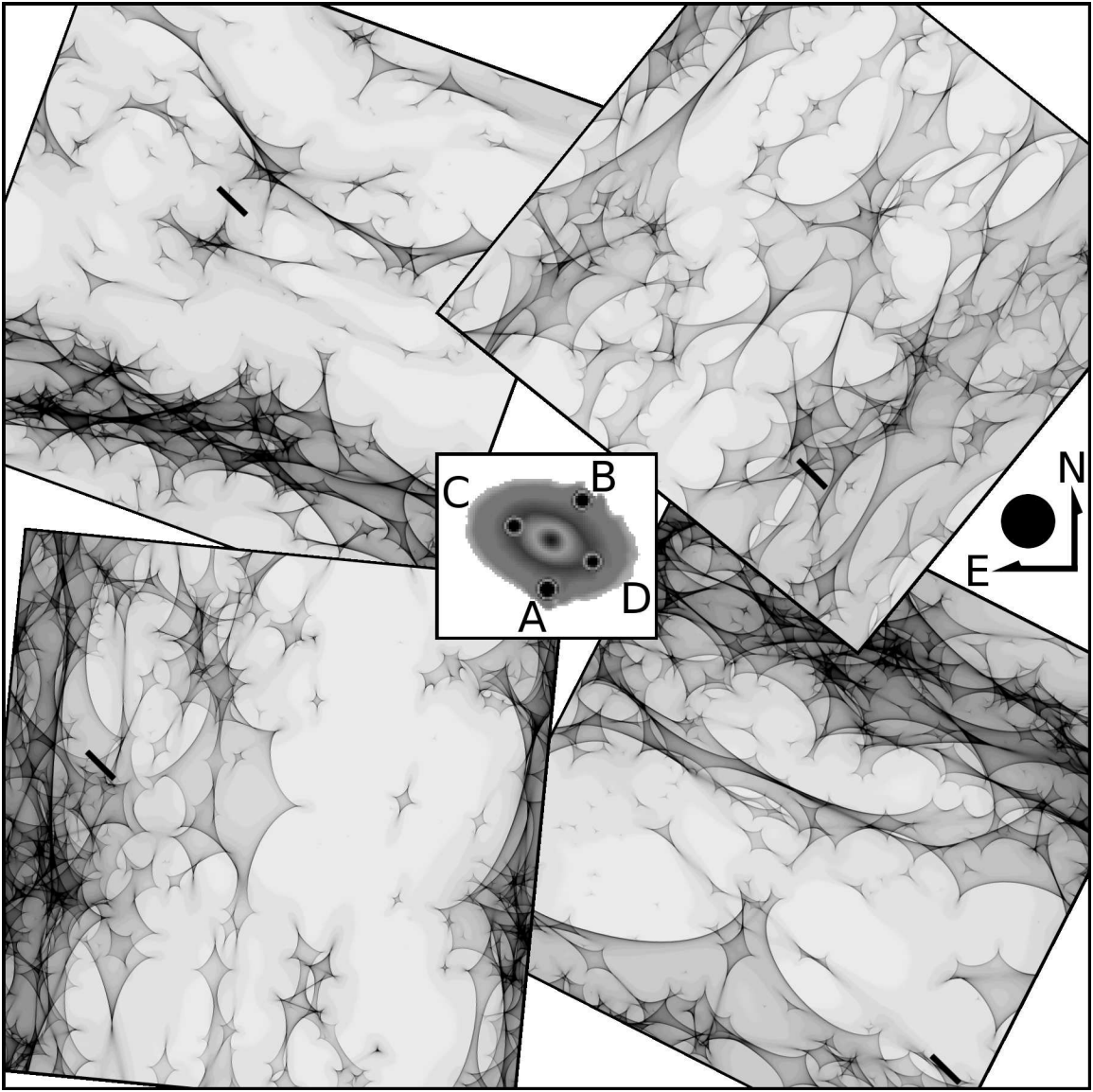}
\caption{Example of a trial source trajectory (dark line segments) superposed on
an instantaneous point-source magnification pattern for $\mmass = 0.3$.
Darker shades indicate higher magnification.
An HST H-band image in the center labels the images and the corresponding
magnification patterns.
Each pattern is rotated to have the correct orientation relative to the lens.
This particular LC2 trial has an effective lens-plane velocity of
$\sim 600\kms$ Northeast.
The solid disk at right has a radius of $10^{17}\cm$ on these patterns.}
\label{fig:tracks}
\end{figure*}

In this paper we use microlensing to measure the peculiar velocity of a lens
galaxy and the mean mass of its stars including the effects of the stellar motions,
Earth's motion, and the rotation of the lens galaxy.
The transverse velocity direction can be measured with microlensing
because the shear sets a preferred direction for each image and the
statistics of variability depend on the motion relative to this axis
(see Figure \ref{fig:tracks}).
In theory, accurately measuring the transverse peculiar velocity
of many galaxies over a broad range of redshifts could form the basis
of a new cosmological test \citep{Gould95}.
Measuring the mean stellar mass, including remnants, is an independent means of
checking local accountings \citep[e.g.][]{Gould00}, which must be assembled from very
disparate selection methods for high mass, low mass, evolved and dead, remnant stars.
Moreover, doing this is possible in detail only for the Galaxy.
While microlensing is relatively insensitive to the mass function
\citep[see][]{pac86,Wyithe00b}, there are some prospects of exploring
this in the future as well \citep[e.g.][]{Wyithe01,Schechter04,Congdon07}.

This work expands on the methods described in \citet{Kochanek04a} and
\citet{Kochanek07} by adding the random stellar motions in the lens galaxy.
In this paper we address the computational issues
and then apply this improved technique to determine the transverse motion of
Q2337 and the mean mass of its stars.
In Paper II,
we study the shape of the accretion disk of the source quasar.
We describe the photometric data in \S\ref{sec:data}.
Then we describe the Bayesian Monte Carlo Method and the models
we use in \S\ref{sec:methods}.
Our results are presented in \S\ref{sec:results} followed by a
discussion in \S\ref{sec:discussion}.
We use an $\Omega_M = 0.3, \Omega_{\Lambda} = 0.7$ flat cosmological model with
$H_0 = 72~\rm km~s^{-1}~Mpc^{-1}$.

\section{Data} \label{sec:data}

The quadruply lensed $z_{\rm s}=1.695$ quasar Q2237 was discovered
by \citet{Huchra85}.
The images are observed through the bulge of a barred Sab lens galaxy
at a projected distance $\sim 0\farcs9$ ($\sim 700$ pc).
The very low $z_{\rm l}=0.0394$ lens redshift leads to very fast lens motions projected onto
the source plane, leading to variability timescales as short as
$\sim 0.2$ years (Equations \ref{eqn:te} and \ref{eqn:ts}).
Microlensing of Q2237 was first observed by \citet{Irwin89} and confirmed by \citet{Corrigan91}.
There are also detailed mass models and dynamical studies by
\citet{Schneider88}, \citet{Kent88}, \citet{Rix92}, \citet{Milhov01},
\citet{Trott02}, and \citet{vandeVen08}.

\begin{figure}
\epsfxsize=3.5truein
\epsfbox{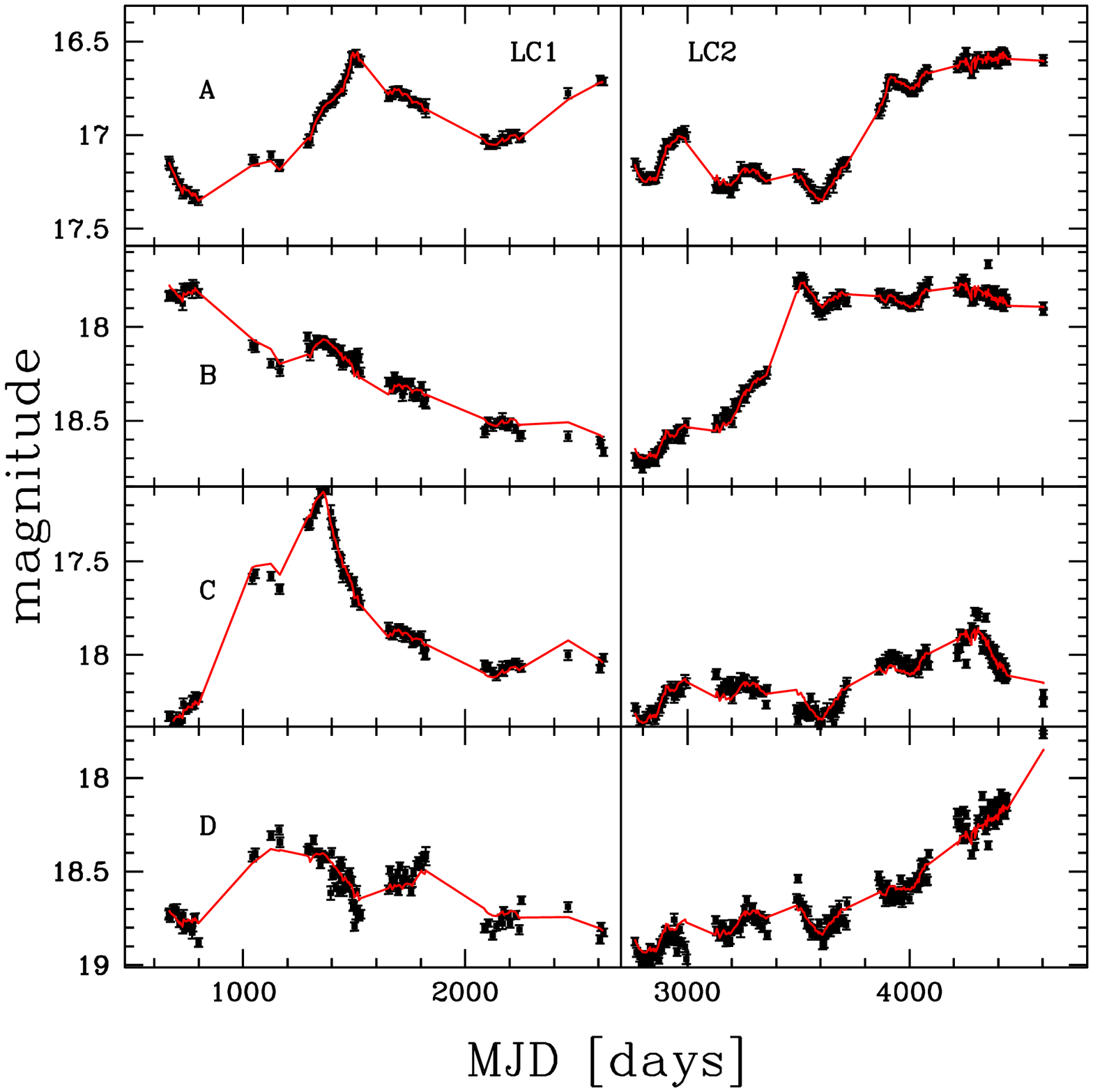}
\caption{The OGLE Q2237 V-band light curves.  The left panel is LC1 and the right is LC2.  The rows
from top to bottom are images A, B, C, and D.  Here we show the corrected error bars.
The red curves are one of our best fit models for the microlensing and intrinsic source
variation.  Because we only determine the light curve at the epochs with data, gaps are filled
by linear interpolation.
\label{fig:lc}}
\end{figure}

We analyze nearly 11 years of the Optical Gravitational Lensing Experiment
V-band photometric monitoring data for Q2237 \citep{Udalski06}.
To speed our analysis and as a cross check on the results,
we divided the OGLE data into two separate light curves.
The first light curve is from JD 2,450,663 to JD 2,452,621 and consists
of 100 epochs and will be referred to as LC1.
The second light curve has 230 epochs from JD 2,452,763 to
JD 2,454,602 and will be referred to as LC2.  Each light curve covers just over 5 years.
The light curves are shown in Figure \ref{fig:lc}.

Since Q2237 is expected to have a very small time delay between its
images \citep[e.g.][]{Wambsganss94}, we only need to subtract the light curves
(in magnitudes) to remove the intrinsic variability of the quasar.
We estimated the systematic photometric errors in the OGLE data using each successive
triplet of epochs spanning less than 15 days.
We used the first and last point of each triplet to predict the middle observation
and then derived the systematic error that, when added in quadrature to the OGLE
uncertainties, make the predictions consistent with the uncertainties.
These systematic error estimates are $0.02$, $0.03$, $0.04$, and $0.05$
magnitudes for images A, B, C, and D respectively.

\section{Methods} \label{sec:methods}

Our Bayesian Monte Carlo method \citep{Kochanek04a} requires the construction
of magnification patterns and a model of the quasar accretion disk.
The patterns are convolved with the source model and used to produce large numbers of
simulated light curves for comparison to the data.
The results for these trial light curves are combined in a Bayesian analysis
to measure parameters and their uncertainties.  Here we describe the generation of the
patterns (\S\ref{sec:magpatterns}), the source model (\S\ref{sec:disk}),
the Bayesian Monte Carlo method and its priors (\S\ref{sec:method}),
and the computational techniques needed to allow for stellar motion (\S\ref{sec:compu}).

\subsection{Magnification Patterns}
\label{sec:magpatterns}

\begin{deluxetable}{cccr}
\tablecolumns{4} \tablewidth{\linewidth}
\tablecaption{Lens galaxy model parameters.\label{tab:lensmodel}}
\tablehead{
  \colhead{Image} & \colhead{$\kappa$} & \colhead{$\gamma$} & \colhead{PA [deg]} }
\startdata
A & 0.40 & 0.40 &  $ 175$ \\
B & 0.38 & 0.39 &  $-39$  \\
C & 0.73 & 0.72 &  $ 70$  \\
D & 0.62 & 0.62 &  $-63$
\enddata
\tablecomments{The normalized surface density ($\kappa$),
shear ($\gamma$) and its position angle at the location at each image.}
\end{deluxetable}

We generate dynamic magnification patterns (see Figure \ref{fig:tracks}
for examples of instantaneous patterns)
in the source plane by randomly placing stars
near each macro image in the lens galaxy.
The normalized surface density and shear are determined by fitting models to the
{\it HST} astrometry of the four images relative to the lens galaxy and the mid-IR image
flux ratios \citep{Agol00}.
We modeled the lens as a power law mass profile with the {\tt lensmodel} program
of the {\tt gravlens} package \citep{Keeton01}.
Because all the images are $\sim 700$ pc in projected distance from the galactic center,
we expect the surface density to be dominated by the stars rather than by dark matter
($\kappa_*/\kappa = 1$).
This assumption is corroborated by the microlensing analysis of \citet{Kochanek04a}
and the dynamical models of \citet{vandeVen08}.
The normalized surface density, $\kappa$, and tidal shear, $\gamma$, from this
model (see Table \ref{tab:lensmodel}) were used to generate the magnification patterns.
The lens plane is populated using a mass function of
$dN/dM \propto M^{-1.3}$ with a dynamic range of $M_{\rm max}/M_{\rm min} = 50$
based on the Galactic mass function of \citet{Gould00}.
We use patterns with mean masses of $\left<M\right>=$
0.01, 0.03, 0.1, 0.3, 1, 3.0, and 10 $M_\sun$.
We use the \citet{Kochanek04a} particle-particle/particle-mesh implementation of the
inverse ray-shooting method \citep[e.g.][]{Kayser86} to create
$N^2_{\rm pix} = 4096^2$ magnification patterns.
The shear is slightly adjusted (by at most $1/N_{\rm pix}$)
in order to produce periodic magnification patterns that eliminate
edge effects, as detailed in the Appendix of \citet{Kochanek04a}.
The outer scale of the patterns is
$20 \rE = 3.7\times10^{18}\mmass^{1/2}\cm$,
which results in a resolution of
$0.005 \rE/{\rm pixel} = 9.0\times10^{14}\mmass^{1/2}$ cm/pixel.
For comparison, \citet{Morgan10} estimate that the black hole mass corresponds
to a gravitational radius of $r_{\rm g} = GM_{\rm BH}/c^2 = 2\times10^{14}\cm$
and in Paper II we find a disk scale length of $6\times10^{15}\cm$
(half-light radius of $1.5\times10^{16}\cm$.

For the first time in any model of microlensing data,
we fully include the random motions of the microlenses by
using an animated sequence of magnification patterns.
We use the measured one-dimensional velocity dispersion of $170\kms$
(van de Ven 2009, personal communication, also \citet{Trott08,Foltz92}).
We assign each star a random velocity as a Gaussian random deviate of
amplitude $\sigma_*=170\kms$ for each coordinate and then
generate a magnification pattern for each image/epoch combination.
While binary stars make up a large fraction of stellar systems \citep[e.g.][]{Fischer92},
they should not have a significant effect on the patterns.
Only relatively close binaries ($\ll 1$ AU) have significant orbital velocities
compared to the stellar or bulk motions, and such separations are very small compared to
the Einstein radius (1100 AU for $1 M_\sun$ in the lens plane).
Thus, binary motion is only significant for close binaries, but close
binaries have separations much smaller than the Einstein radius or our
estimated source sizes (66 AU for $1 M_\sun$ in the lens plane, see Paper II) and
would be indistinguishable from a single point mass on our patterns.
In effect, binaries should only act like a shift in the mean mass of the stars.

\subsection{Disk Model}
\label{sec:disk}

We employ a generic thin disk model for which the surface temperature
scales as $T_{\rm s} \propto R^{-3/4}$ with radius R \citep{Shakura73}.
The microlensing signal is primarily sensitive to the half-light radius of the
disk \citep{Mortonson05}, so the details of the radial profile have limited effects.
As in \citet{Kochanek04a}, we neglect the inner disk edge, since it should have
few observed effects given disk sizes at these wavelengths.
We define the area of the disk to be the area enclosed within the contour defined by
$kT = hc/\lambda$.
We first parametrize the source models by choosing from 24 different projected
areas covering $\log_{10}({\rm area/cm}^2) = 29.2$ to $33.8$ in steps of $0.2$.
For each source area, we used five inclinations, $i$, with a $\cos i$ of
0.2, 0.4, 0.6, 0.8, and 1.0 (face-on), and for each area and inclination, we used
18 different equally spaced position angles for the major axis of the disk.
Paper II discusses the disk model in detail.

\subsection{The Bayesian Monte Carlo Method}
\label{sec:method}

We use the Bayesian Monte Carlo Method of \citet{Kochanek04a}.
We randomly generate light curves from the animated microlensing magnification
patterns over the full range of physical parameters and source sizes
and fit them to the observed light curves.  We then use Bayes Theorem
to infer the likelihood distribution of the parameters given the fit
statistics for the light curves.
Each simulated light curve is defined by
\begin{equation}
m_i(t) = S(t) + \mu_i + \delta\mu_i(t) + \Delta\mu_i
= S(t) + \mu_{i,\rm tot},
\label{eqn:lc}
\end{equation}
where $S(t)$ is the intrinsic variability of the source,
$\mu_i$ is the macro model magnification,
$\delta\mu_i(t)$ is the microlensing, and
$\Delta\mu_i$ is the systematic magnification offset
for each image, $i$.
For each trial we compute the goodness of fit
\begin{equation}
\chi^2 = \sum_i\sum_t \left(\frac{m_i(t) - S(t) - \mu_{i,\rm tot}}{\sigma_i(t)}\right)
\end{equation}
after solving for the optimal model of the source variability $S(t)$ and the
magnification offsets.  The parameters we vary in this study
include the projected area of the disk, the inclination of the disk, the position angle
of the disk, the effective velocity of the source, and the mean mass of the stars
in the lens galaxy.  We call these the physical parameters, $\xip$.
For any combination of these parameters, we also randomly select starting points
on each of the magnification patterns and refer to these
nuisance parameters, $\xin$.

We calculate the likelihood of the data given the parameters as
\begin{equation}
P(LC|\xip,\xin) \propto \Gamma\left[\frac{N_{\rm dof}}{2},\frac{\chi^2}{2}\right],
\label{eqn:gamma}
\end{equation}
where $\Gamma$ is the incomplete Gamma function. \citet{Kochanek04a} justifies
this form by allowing for uncertainties in the magnitude errors, $\sigma_i(t)$
and averaging over these uncertainties.  
This ensures that the likelihood is consistent for trials fitting better than
$\chi^2/{\rm dof} = 1$ given the formal uncertainties.

The probability of the parameters given the data is then
\begin{equation}
P(\xip,\xin|LC) \propto P(LC|\xip,\xin)P(\xip)P(\xin),
\end{equation}
where $P(\xip)$ and $P(\xin) = 1$ are the prior probability distributions
of the physical and nuisance parameters.
Since we are analyzing two separate parts of the same light curves, we
combine the results to improve our measurements by multiplying the
probabilities for each light curve and then applying the priors,
\begin{equation}
P(\xip|LC1,LC2) \propto P(LC1|\xip)P(LC2|\xip)P(\xip).
\end{equation}
We did this for two reasons.
First, it becomes (probably exponentially) harder to find good fits to longer
and longer light curves.
Second, analyzing the curves separately allows us to study whether
different light curves for the same object give the same answers.
The price is that analyzing them separately and then combining them will have
less statistical power than a simultaneous analysis of all the data.
We compute the probability distributions by marginalizing over
the nuisance variables
\begin{equation}
P(\xip|LC) \propto \int P(\xip,\xin|LC)d\xin.
\label{eqn:monte}
\end{equation}
We compute this as a Monte Carlo integration over the trial light curves,
which should converge to the true integral if we generate enough simulated light curves.

For each source size, inclination, and disk position angle we must first
convolve the magnification pattern with the source model.  Then we produce trial
light curves by choosing random starting points and velocities across the
animated sequence of magnification patterns.
In addition to the random motions of the stars, we must also assign bulk velocities
to the observer, $\vvo$, lens, $\vvl$, and source, $\vvs$, leading to an
effective (source-plane) velocity of
\begin{equation}
{\bf v}_{\rm e} =
  \frac{\vvo}{1+z_{\rm l}}\frac{\dls}{\dol}
- \frac{\vvl}{1+z_{\rm l}}\frac{\dos}{\dol}
+ \frac{\vvs}{1+z_{\rm s}}
\end{equation}
\citep[e.g.][]{Kayser86}
that is dominated by the lens velocity, ${\bf v}_{\rm l}$, in the case of Q2237.
From the projection of the CMB dipole \citep{Hinshaw09}, we know that $\vvo=(-50,-23)\kms$
East and North respectively for Q2337. Based on
\citet{Tinker09} we estimate that the (1D) rms peculiar velocities of the lens and source
are $\sigma_{\rm lens}=327\kms$ and $\sigma_{\rm src}=230\kms$, respectively.
For the calculations, we randomly draw each effective velocity coordinate (in the lens plane)
from a one dimensional Gaussian distribution with $\sigma=1000\kms$ and then re-weight
the trials to a more physical range when we carry out the Bayesian integrals.

As discussed in the introduction, quasar microlensing is subject to a
degeneracy between mean stellar mass, effective velocity, and accretion disk size.
Including random stellar motions partially breaks this degeneracy by
introducing a physical scale to the magnification patterns.
We still need a prior on one of these parameters to make useful measurements.
As in \citet{Kochanek04a}, we apply a velocity prior. Here we define our prior in
the lens-plane, since the lens motion dominates the effective velocity.
In the absence of any ``streaming velocities'', we can determine the peculiar velocities
only up to a $180\deg$ degeneracy that corresponds to a time reversal symmetry given
that the peculiar velocity priors depend on the speed but not the direction of motion.
Our velocity prior in the lens-plane is
\begin{eqnarray}
P(\vvl) &\propto& \exp\left(-\frac{(\vvl-\vvc)^2}{2\sigma^2}\right),
\label{eqn:vprior}
\end{eqnarray}
where $\vvc$ is our CMB motion projected onto the lens-plane, and
the expected dispersion in the lens-plane is
\begin{eqnarray}
\sigma^2 &=& \sigma_{\rm lens}^2 + (\sigma_{\rm src} (1+z_{\rm l}) D_{\rm OL}/D_{\rm OS})^2 \nonumber \\
&=& (327\kms)^2.
\label{eqn:sigma}
\end{eqnarray}
The very high projected motion of the lens due to the very low lens redshift
means that the source motion is unimportant
even though $\sigma_{\rm src} \sim \sigma_{\rm lens}$.
``Streaming velocities'', such as our motion relative to the CMB, our orbit around
the Earth (parallax effect), or rotational velocities in the lens break this degeneracy.
These motions (up to $\sim 10\%$ of the peculiar velocties), do slightly break this degeneracy.

With the stars moving it is also makes sense to include the
effects of the Earth's motion and the small rotation velocity
of the lens galaxy as part of the motions across the animated
patterns.  Aside from \citet{Tuntsov04} the motion of the
Earth has not previously been included in a quasar microlensing calculation.
Earth's motion projected onto the lens plane is approximately
10\% of the expected transverse velocity of the lens motion (the dominant motion of
the system).  It is also $\sim 20\%$ of the minimum possible velocity scale set by
the random stellar motions.
In trials with $\mmass = 0.3$, we found that including parallax increased the total probability
of all trials by $\sim 20\%$, and reversing the Earth's motion reduced the probability
by a similar amount.
Earth's orbit is trivial to include and computationally inexpensive, so we include it
in our standard calculations even though it has modest effects on the likelihood.

The lens is an late type galaxy with rotation in the plane of the sky of $\sim 55\kms$ for
images A and B, and $\sim 20\kms$ for images C and D \citep{Trott08}.
The position angle of these rotation velocities are $84.5\deg$, $-129\deg$, $-20\deg$,
and $153\deg$ (north through east) respectively for images A, B, C, and D.
These are relatively low compared to the disk because the images lie in the bulge.
The velocities for
images A and B are greater than the modulations introduced by the Earth's orbit,
so we include them in the simulation.

\subsection{Computational Techniques}
\label{sec:compu}

The Monte Carlo method requires simultaneous random access to every
magnification pattern for each image and epoch.
With the stars moving, this means we need 400 and 920
patterns for LC1 and LC2 respectively, corresponding to 25 and 57.5 gigabytes
of storage for $4096^2$ patterns instead of the 4 patterns and 0.25 gigabytes
needed for stationary stars.
This is more memory than is generally available on any one machine.

Our first step towards solving this computational problem is to conserve memory by compressing
the ``gray'' scale of the convolved patterns.
Normally we store patterns as a $4096^2$ array of 4-byte floating point numbers.
However, magnification patterns typically span a dynamic range of $10$ magnitudes
magnitudes even for the smallest source sizes, and the data uncertainties
are no smaller than 0.01 magnitudes.
Thus we only need a logarithmic dynamic range of $10/0.01 = 1000$
rather than the $2^{32}$ dynamic range of a floating point variable.
For example, if we use 16 bits for each pixel, which is more than sufficient given
the dynamic range of the data, we can pack 4 pixels into one 64 bit word rather than using
128 bits, which not only compresses the data by a factor of 2 but also has advantages
for data transfer speeds.
In practice, we adjust the compression level for each magnification pattern
and source size to have a resolution at least ten times better than the
uncertainty in the corresponding data point.
We achieve compression ratios of 2.5 to 3 for the OGLE data.

Even after compression, the full collection of magnification patterns is still too large
for most single machines, so we distribute them evenly among parallel computers.
This has the added benefit of utilizing additional CPUs, but at the cost
of needing to communicate between nodes.
Our goal is to minimize the need for this communication.
We sort each light curve in chronological order and then distribute the epochs in a
round robin fashion to each node, so that each node has a sparse but
complete representation of the data. Trial light curves are started on the
individual nodes.  If a trial's $\chi^2$ exceeds a threshold it is simply discarded.
If it is under the threshold, it is passed to other nodes to be tested against
the rest of the light curve.
This basically amounts to a low resolution pre-search for good-fitting light curves
before doing any expensive communication with the other nodes.
These light curves are optimized by exploring slightly different
starting points and velocities across the magnification patterns.
This requires the master node for each trial to do many communications
with the other nodes to compute the full $\chi^2$, but our tests show that this
finds good fits faster than trying more light curves.

For each source model, we choose $10^5$ starting points and velocities for one
of the four images.  Then we search for pairwise matches by trying $10^4$ starting
points on each of the other three images and keep the best match for each image.
A light curve is then produced from this velocity and starting point for
each image.  The $\chi^2$ for each trial light curve is computed from the data.  
If the $\chi^2$ of a trial exceeds a threshold during its calculation, we
discard it immediately.  Such poor solutions will make no contribution to the
Bayesian integrals (Equation \ref{eqn:monte}), so there is no point in wasting
further calculation or communication on completing the trial.

LC1 was processed in 1.6 CPU-years and LC2 was processed in 2.8 CPU-years utilizing
16 AMD Opteron machines (64 processor cores) simultaneously at the Ohio Supercomputing Center.
In total we tried $9\times10^{14}$ unique starting points and $3\times 10^{9}$
different velocities.
We found significantly fewer good fitting trials for LC1, so our threshold
for saving trials was $\chi^2/{\rm dof} < 4$ for LC1 and $\chi^2/{\rm dof} < 2.5$ for LC2.
Our best fit simulated light curves have $\chi^2/{\rm dof} = 0.86$ for LC1,
and $\chi^2/{\rm dof} = 0.99$ for LC2.
There was a large event in image C, and more rapid magnification changes
in LC1 as compared to LC2 (see Figure \ref{fig:lc}), and this likely
explains why it was harder to find good fits for LC1.
With these cuts, $3\times 10^6$ and $6\times 10^6$ trials fits passed the cuts for LC1
and LC2 and were saved.
Even though the best fitting light curves produced $\chi^2/{\rm dof}\sim 1$,
we rescaled the $\chi^2$ for each light curve to produce better-defined results
in the Monte Carlo integral (Equation \ref{eqn:monte}) for each parameter.
We divided the $\chi^2$ of trials of LC1 and LC2 by
\lcOneRescale and \lcTwoRescale respectively for our final analysis,
so that $10^4$ trials for each set were less than $\chi^2/{\rm dof}$ after rescaling.
In general, this is conservative and broadens the parameter uncertainties
by $\sqrt{\lcOneRescale}$
and $\sqrt{\lcTwoRescale}$ over what we should achieve with an infinite number of trials.
 
\section{Results} \label{sec:results}

\begin{figure}
\epsfxsize=3.5truein
\epsfbox{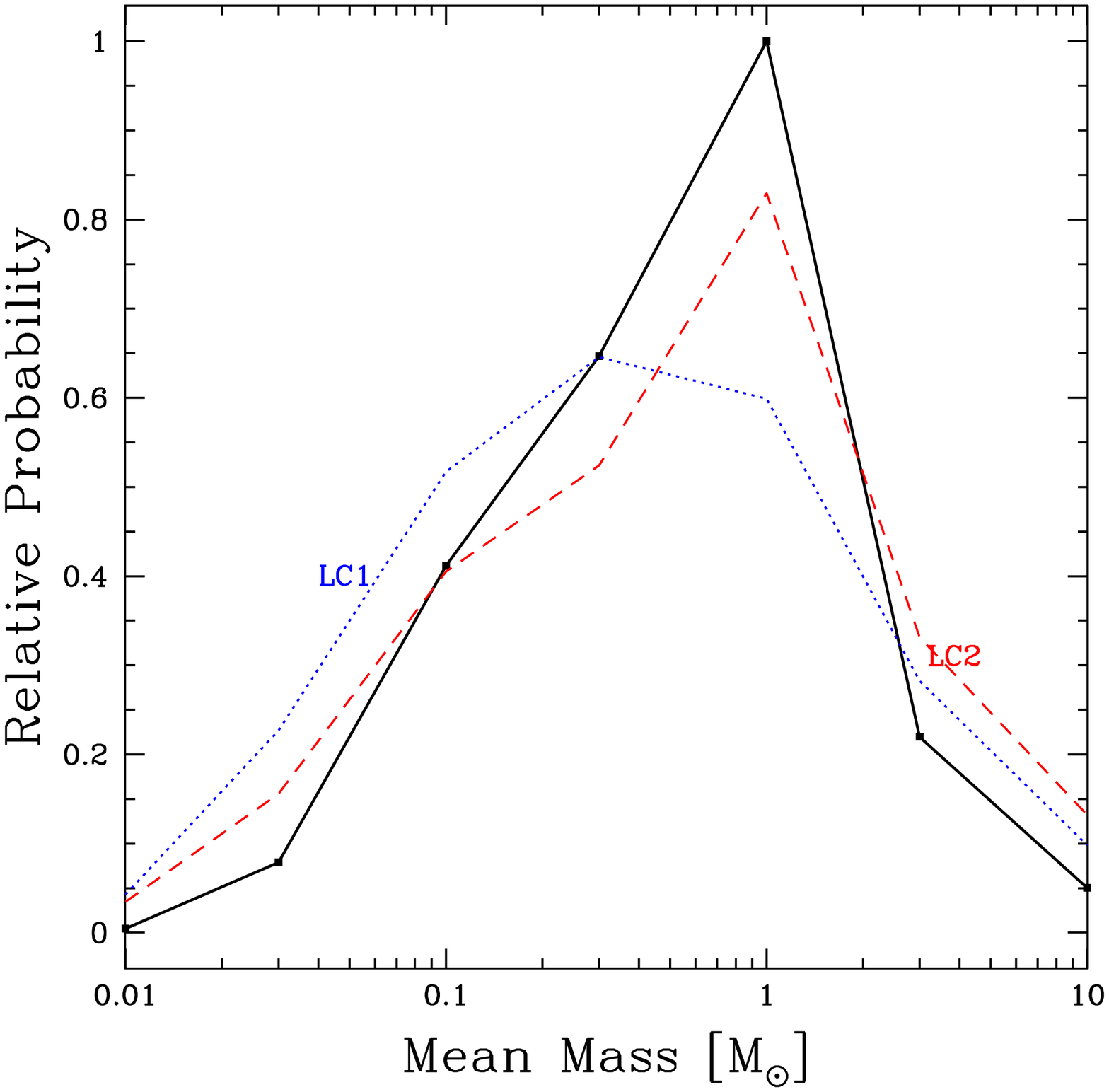}
\caption{Mean mass of the stars. \bluered
\label{fig:mass}}
\end{figure}

Here we estimate the transverse peculiar velocity of the lens galaxy,
the mean mass of its stars, and the mean magnification offsets defined
in Equation \ref{eqn:lc}.
We quote the results from the combined analysis of LC1 and LC2,
but also show the results from the independent analyzes of LC1 and LC2 in Figures
\ref{fig:vel}, \ref{fig:vang}, \ref{fig:speed}, and \ref{fig:mass}.
Our main results are based on the $\sigma=327\kms$ lens-plane (1D) peculiar velocity prior
described in \S\ref{sec:method} and assume that all the light comes directly from the
accretion disk.  We verify the latter assumption and
examine the structure and orientation of the accretion disk in Paper II.

\subsection{Transverse velocity}

\begin{figure}
\epsfxsize=3.5truein
\epsfbox{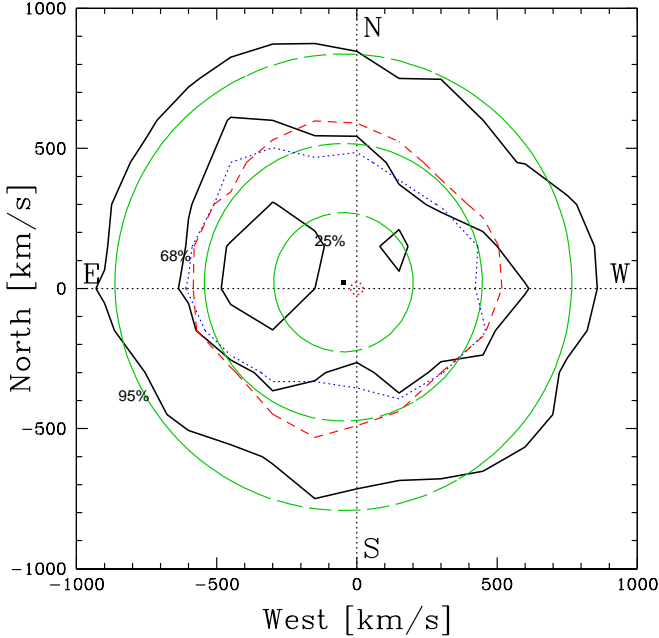}
\caption{
Probability distribution for the effective lens-plane velocity.
The dark solid contours enclose 25\%, 68\%, and 95\% of the likelihood relative
to the peak.
Because the motion is strongly dominated by the lens, this is essentially the
transverse peculiar velocity of the lens galaxy.
The green dashed contours are the velocity prior (Equation \ref{eqn:vprior}),
drawn at the same levels.
The small, red, dotted circle is the 68\% contour for the contribution to the
prior from the expected motion of the source.
The black point is our CMB motion projected onto the lens-plane.
The 68\% enclosed probability for the LC1 and LC2 analyses are shown
as dotted blue and dashed red contours.
\label{fig:vel}}
\end{figure}

\begin{figure}
\epsfxsize=3.5truein
\epsfbox{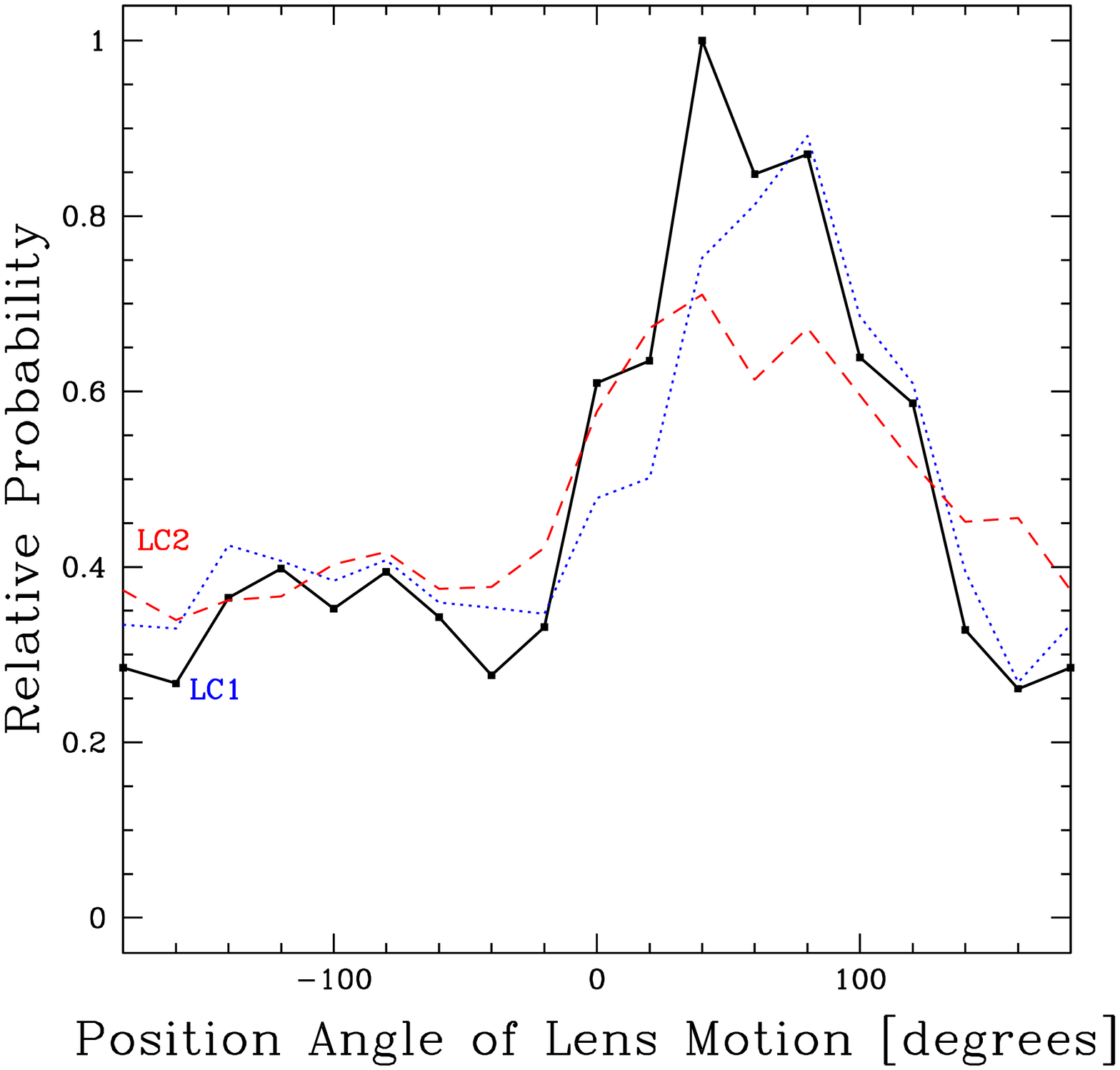}
\caption{
Position angle of the effective velocity in the lens-plane.
\bluered
\label{fig:vang}}
\end{figure}

\begin{figure}
\epsfxsize=3.5truein
\epsfbox{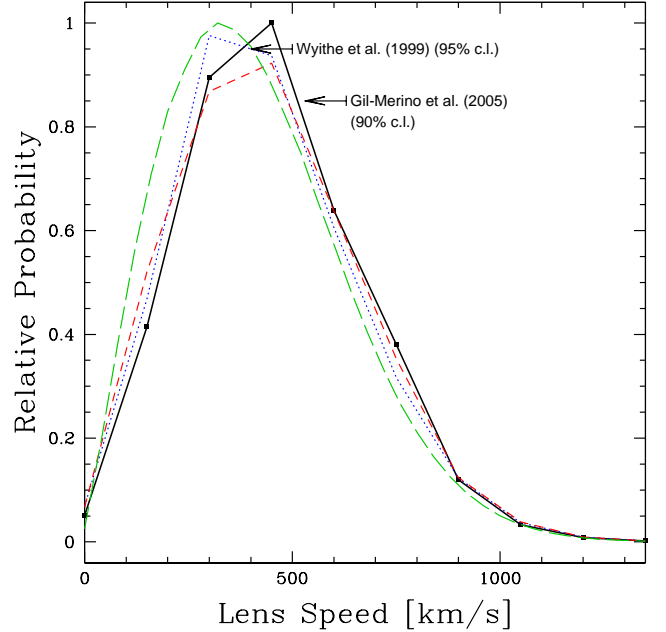}
\caption{Speed distribution of the effective velocity of the source in the lens-plane 
(dark solid curve).  This motion is dominated by the priors on the transverse
peculiar velocity of the lens.
The green dashed curve shows the prior (Equation \ref{eqn:vprior}) integrated over direction.
The upper limits from other studies are indicated with arrows \citep{Wyithe99,GilMerino05}.
\bluered
\label{fig:speed}}
\end{figure}

Figure \ref{fig:vel} shows the likelihood contours for the transverse
velocity of the lens galaxy.
The peak likelihood is $\sim 300\kms$ East.
The individual light curve results are very consistent with the joint analysis, as
shown by superposing the 68\% contours for LC1 and LC2 in Figure \ref{fig:vel}.
This agreement can also be seen in the position angle of the lens motion
(Figure \ref{fig:vang}) and in the lens speed (Figure \ref{fig:speed}).

After integrating over direction we find a transverse speed of
$438^{+253}_{-213}\kms$ ($438^{+419}_{-325}\kms$)
at 68\% (95\%) confidence (Figure \ref{fig:speed}) including our standard prior
(Equations \ref{eqn:vprior} and \ref{eqn:sigma}).
The inclusion of a physical model of the stellar motions does not completely eliminate
degeneracies, and our speed estimate is dominated by our velocity prior
(Equation \ref{eqn:vprior}) at large speeds.
If we instead use the broader lens-plane prior with $\sigma=1000\kms$
from which we derive our trials,
the measured speed becomes $1048^{+640}_{-486}\kms$ at 68\% confidence basically
following the prior.
Therefore, we only have measured a lower limit $v_{\rm t} > 338\kms$.
Fixing the mean microlensing mass has little effect, provided the mass is
sufficiently large.
If we fix the mean mass to be $\left<M\right> = 0.3 M_\sun$ we find that
$v_{\rm t} < 486~(757) \kms$ at 68\% (95\%) confidence.

These results are consistent with earlier results.
\citet{Wyithe99} found a 95\% upper limit on the transverse lens speed,
$v_{\rm t} < 500\kms$ by using derivatives of the microlensing light curve from patterns
produced from a Salpeter mass function including stellar motions and a range of source sizes.
They tried three different mass models with $\left<M/M_\sun\right> = 0.22,0.31$ and $1.0$
and found that the estimated transverse speed scaled with $\sqrt{\left<M\right>}$.
Using the distribution of gaps between high magnification events, but not including
stellar motions, \citet{GilMerino05} found a 90\% upper limit, $v_{\rm t} < 630 (2160) \kms$
for $M=0.1 (1.0) M_\sun$ lenses.
It must be noted that neither study included the full range of physical uncertainties
we include here.

\subsection{Mass}
\label{sec:mass}

\begin{figure}
\epsfxsize=3.5truein
\epsfbox{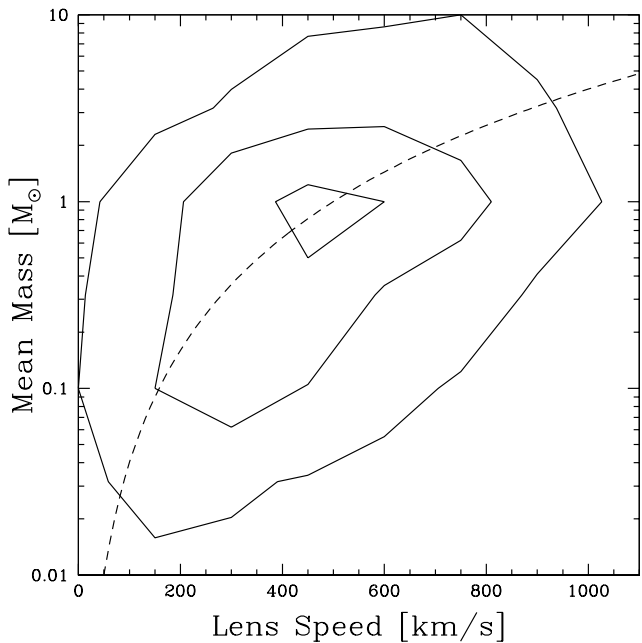}
\caption{Mean mass, $\left<M\right>$ of the stars versus the speed of the lens, $v$.
The dashed line is $M \propto v^2$.
\label{fig:speedmass}}
\end{figure}

We measure the mean stellar mass in the lens to be
$0.12 < \mmass < 1.94$ ($0.04 < \mmass < 3.46$)
at 68\% (95\%) confidence with a median of $\mmass = 0.52$.
This is generally consistent with earlier estimates for this lens based
on less data and simpler analyses including fewer of the physical uncertainties.
It is marginally consistent with the earlier estimate by
\citet{Kochanek04a} of $\mmass = {0.018}^{+0.080}_{-0.015}$
(${0.018}^{+0.270}_{-0.017}$) at a 68\% (95\%) confidence using
a similar velocity prior but with less data and static stars.
Our uncertainties are a factor $\sim 2$ times smaller.
\citet{Lewis96} argue for a mean mass of $0.1 < M/M_\sun < 10$
by comparing the observed magnification probability distribution to that
of simulations that did not include random stellar motions.
\citet{Wyithe00b} estimate that $\mmass = 0.29$ with a lower limit of
$\mmass \ga 0.11$ at 99\% confidence
by analyzing the distribution of light curve derivatives and
assuming a 1D stellar dispersion of $165\kms$ combined with a prior on the
transverse velocity.
\citet{GilMerino05b} simply argued that the masses must be greater
than Jupiter-like objects, contrary to the claims of \citet{Lee05}.
In our results, 
the mass estimate is still strongly affected by our velocity prior
(Equation \ref{eqn:vprior}).
If we use the broad $\sigma=1000\kms$ lens-plane velocity prior,
the median rises to $\mmass = 1.5$, and we would need to expand
the calculations to higher masses to fully sample the mass distribution.
The problem for accurate mass measurements is that $\left<M\right> \propto v^2$
(see Figure \ref{fig:speedmass}),
so the mean mass is very sensitive to the speed distribution (see Figure \ref{fig:speedmass}).
Including the mean stellar motions eliminates low velocity solutions
corresponding to low masses, but cannot eliminate high mass, high velocity
solutions.

\begin{figure}
\epsfxsize=3.5truein
\epsfbox{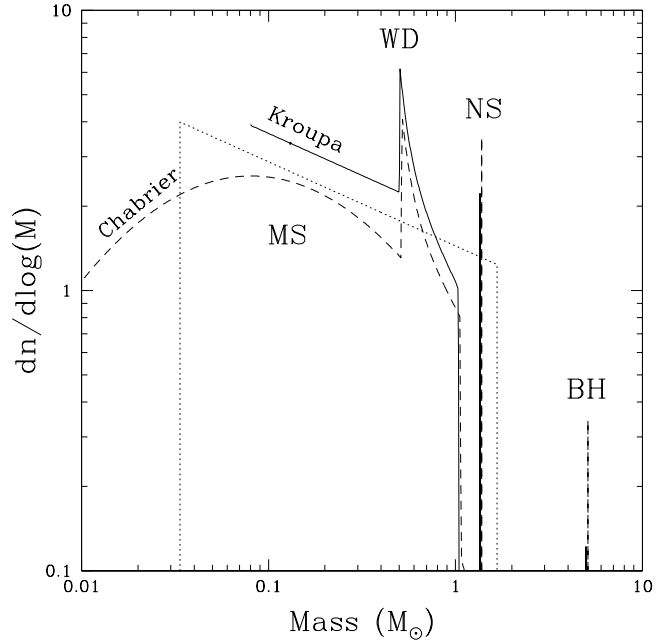}
\caption{
The most probable $\langle M \rangle =0.3M_\odot$ microlensing mass
function used in the calculations (dotted curve) as
compared to that predicted from the
\citet{Chabrier03} or \citet{Kroupa93} initial mass functions truncated at
masses lower than $0.01M_\odot$ (mean mass $0.20M_\odot$) and
$0.08M_\odot$ (mean mass $0.32M_\odot$), respectively, for an age of
$10$~Gyr. The features due to the main sequence (MS), white dwarfs (WD),
neutron stars (NS) and black holes (BH) are marked.
The \citet{Chabrier03} and \citet{Kroupa93} mass functions are slightly
offset to make the different amplitudes for neutron stars and black holes visible.
The assignment of a $5M_\odot$ mass for black holes is arbitrary but not important.
\label{fig:massfn}}
\end{figure}

We can compare this mass estimate to that expected from stellar
mass functions.  We approximate the lifetimes of stars by the time
to reach the base of the giant branch, using the approximate expression
in \citet{Hurley00}.  The remaining lifetimes beyond this point
are an unimportant correction.  Stars older than this lifetime are
modeled as remnants using the white dwarf initial/final mass relation
$M_{\rm WD}=0.109 M +0.394 M_\odot$ of \citet{Kalirai08}
for $M_{\rm NS} < 8M_\odot$, a neutron star
mass of $1.35 M_\odot$ for $8M_\odot < M < 40M_\odot$, and a black
hole mass of $5 M_\odot$ for masses $M_{\rm BH} > 40 M_\odot$.
Using the initial mass function from \citet{Chabrier03}, a combination
of a log-normal distribution at low mass and a power-law at high
mass covering $0.01 M_\odot < M < 100 M_\odot$, the mean mass is
\begin{equation}
 \langle M \rangle \simeq
     \left(0.20+0.03\log\left(t/10~\hbox{Gyr}\right)\right) M_\odot
\end{equation}
for any reasonable population age $t$.
Age has little effect because the high mass stars which
evolve on these time scales make a limited contribution to the mean mass,
and the mass scale beyond which stars have evolved depends weakly on age
($M_{evolve} \simeq (t/11~\hbox{Gyr})^{0.31} M_\odot$ for the
\citet{Hurley00} models).  Changes in the white dwarf mass
relations also have little effect. For $M_{WD}=a M+b$, the
sensitivity is
$\delta \langle M \rangle \sim 0.08 \delta a M_\odot + 0.05 \delta b$,
where \citet{Kalirai08} estimate uncertainties of $\delta a \simeq 0.007$
and $\delta b \simeq 0.025 M_\odot$.  Similarly, changes in the
masses of neutron stars and black holes have negligible effects on the mean mass, with
$\delta \langle M \rangle \simeq 0.0035 \Delta M_{\rm NS}$ and
$\simeq 0.00034 \Delta M_{\rm BH}$, respectively, and the same holds
true for the masses defining the boundaries between remnant types.
Even giving all stars a binary companion with the secondary mass ratio
uniformly distributed from $1/50 < M_2/M_1 < 1$ affects the mean mass
little, roughly $0.05 M_\odot$.  The sense of the effect depends on the
size distribution of the binaries.  Very wide binaries act like independent
stars and so lower the mean, while very close binaries act like a single, higher
mass star and hence raise the (effective) mean.

Thus, only changes in the actual initial mass function can significantly
alter the expected mean mass.  The \citet{Chabrier03} mass function converges
to low masses, so extending the mass range downwards to $0.001 M_\odot$ from
$0.01 M_\odot$ reduces the mean mass by only $0.02 M_\odot$.  Significant
changes require a mass function converging more slowly at lower masses or
adding entirely new populations.  For example, if we instead use a
\citet{Kroupa93} mass function, the results are very sensitive to the low mass cutoff
because the mass function is a rising power law ($\propto M^{-1.3}$) to
low masses.  For minimum masses of $0.08M_\odot$ and $0.01 M_\odot$ the mean
masses are $0.32M_\odot$ and $0.15 M_\odot$ respectively.  Figure \ref{fig:massfn}
compares these mass functions to our model simple power law model with
$\langle M\rangle =0.3M_\odot$ and $M_{\rm max}/M_{\rm min} = 50$,
to show that our maximum likelihood model is in good agreement with expectations
for normal stellar populations.

We can also compare this mean mass to microlensing measurements made in
our own Galaxy and in other quasar microlensing studies of other lenses.
The MAssive Compact Halo Object (MACHO) survey measured
the most likely mass range of compact objects in the Milky Way Halo to be
$0.15 < M/M_\sun < 0.9$ depending on the halo model \citet{Alcock00},
although these results are broadly questioned
\citep[e.g. see][]{Tisserand07,Wyrzykowski09}.
Estimates for the Galactic bulge are probably more relevant for comparison
to Q2237.
\citet{Han96} determined that a power law mass function
$dN/dM \propto M^{-2.1}$ for $M > 0.04 M_\odot$
was the best fit to a sample of 51 MACHO Galactic bulge microlensing events.
This corresponds to a mean mass of $0.19M_\sun$ assuming a maximum mass of $10M_\sun$.
\citet{Grenacher99} studied the first 41 MACHO bulge events toward Baade's windows and
found a mean mass of $0.09M_\sun$ ($0.129$) for bulge (disk) lenses.  They assumed
a Salpeter mass function in the range 1-10 $M_\sun$ and fit for the best slope and minimum
mass below $1M_\sun$.
\citet{Calchi08} found a very similar result.
Outside our Galaxy, the only limits aside from those for Q2237 are those for the
doubly imaged quasar Q~0957+561 by \citet{Schmidt98}, who found a weak lower limit of
$\left<M\right> \ga 0.001 M_\odot$.

\subsection{Magnification Offsets}

\begin{figure}
\epsfxsize=3.5truein
\epsfbox{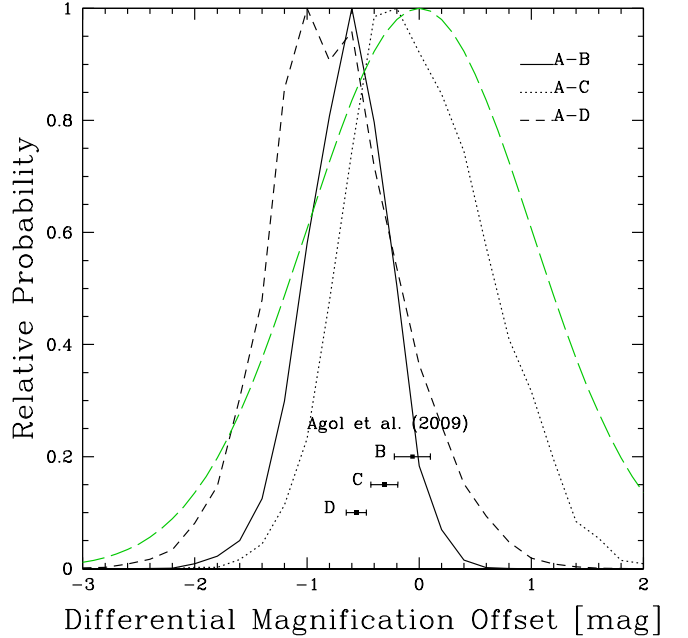}
\caption{Differential magnification offsets, $\Delta\mu_i$ for A$-$B, A$-$C, and A$-$D.
The extinction measurements by \citet{Agol09} are shown relative to image A.
The dashed curve is the prior we used on the magnification offsets.
\label{fig:zero}}
\end{figure}

We can also try to measure the relative mean magnification offsets between each of the images.
In our models we do not constrain the mean magnification ratios of the images
to closely match the predictions of the lens model, since differential dust extinction
\citep[e.g.][]{Falco99,Eigenbrod08a,Agol09},
undetected substructure \citep[e.g.][]{Mao98,Metcalf01,Kochanek04b,Vegetti09},
bad ``macro'' models of the lens magnification, and contamination of the light curves by
light from the lens or host galaxy can also change the relative brightnesses of the images.
We allow $\Delta\mu_i$ (Equation \ref{eqn:lc}) to be optimized for each fit, subject to a
Gaussian prior with a 1.0 magnitude dispersion.
In Figure \ref{fig:zero} we show the posterior probability distributions
for these differential offsets.
For an infinitely long light curve, these offsets will converge to zero
in the absence of any systematic problems.
The differential offsets between A$-$B and A$-$D show weak evidence
for offsets, but there is surprisingly little convergence in their values.
For comparison, \citet{Agol09} used the flux ratios of the quasar broad lines
as compared to the continuum from \citet{Eigenbrod08a}
to estimate the extinction of the images relative to A.
They found $\Delta E(B-V)=0.02\pm0.05, 0.10\pm0.04$, and $0.18\pm0.03$
for images B, C, and D respectively.
Figure \ref{fig:zero} shows these estimates assuming a $R_{\rm V} = 3.1$ extinction curve.
There is some correlation between our estimates and these shifts,
but our estimates are simply too uncertain to draw any conclusions.
We experimented with forcing our trials to match the extinction estimates of
\citet{Agol09} by multiplying the probability of each trial by a Gaussian
model of these extinction estimates.
We found no significant influence on any other parameter distribution.
\citet{Dai09} reached a similar conclusion in their analysis for RXJ1131$-$1231.

\section{Discussion} \label{sec:discussion}

By including the random motions of the stars, we can now use microlensing to
study the peculiar velocity of the lens galaxy and to estimate mean stellar masses
and potentially the stellar mass functions with fewer systematic uncertainties.
In particular, we find a clear preference for the direction of motion of the lens galaxy.
In fact, as we
use a less restrictive velocity prior, the direction of motion is better
constrained since faster speeds are allowed.
We cannot however, determine the speed without additional priors.
If we assume a mean stellar mass of $\left<M\right> = 0.3 M_\sun$,
we find that the peculiar velocity is $v_{\rm t} < 486 \kms$ which is consistent
with the other estimates \citep{Wyithe99,GilMerino05} but more fully
includes all the physical uncertainties.
It should not be surprising that we can determine a dimensionless quantity, the direction
of motion, better than the dimensional speed, given the basic problem of microlensing
that all observables are in $\sqrt{\left<M\right>}\cm$.
Very roughly (see Figure \ref{fig:tracks}), the preferred direction has images A and B
moving more closely to perpendicular to the ridges of the magnification patterns
created by the shear and images C and D are moving more parallel to the shear direction.
This is consistent with the variability of A/B compared to C/D.
We included the parallax effects of the Earth's orbit, and the results weakly
favored its inclusion.

We had hoped that modeling the random motions would be more of a help in breaking these
degeneracies by setting a physical scale.
This is probably true for low mean masses $\left<M\right>$.
For fixed variability amplitudes, reproducing the light curves with a low mean mass
requires small physical velocities, while high mean masses require high velocities.
Adding the stellar motions at their observed dispersion
eliminates low mass solutions independent of the unknown peculiar velocities by setting
a floor to the velocity scale.
High mass solutions need peculiar velocities, $\sigma$, that are larger than the stellar motions,
$\sigma_*$, and so are only constrained by
the priors on the peculiar velocities.
Essentially, the dynamic patterns act like static patterns once
$\sigma_* < \sigma$, and we recover the familiar degeneracies of static patterns.
Thus, our correct treatment of the stellar motions constrains low mass but not high
mass solutions in the absence of a peculiar velocity prior.
With a well-defined cosmological prior on $\sigma$ \citep{Tinker09}, we find that
$0.12 \leq \left<M/M_\sun\right> \leq 1.94$ at $68\%$ confidence,
demonstrating that the microlensing objects are typical of stellar populations and their remnants. 
This mass range is consistent with expectations for normal stellar populations
(see \S\ref{sec:mass}), but not tightly constraining.

We largely ignore the macro magnifications predicted by the mass distribution of
the lens galaxy in our calculations because of their systematic uncertainties.
However some recent studies have made use of this information by analyzing
image pairs straddling a critical curve which should have the same magnification
\citep{Floyd09,Bate08}.
A concern is that the macro magnification may be affected by undetected substructure,
differential extinction, or contamination by the lens or host galaxy.
In our standard analysis we use the AC signal and largely discard the DC
signal by not tightly constraining the mean magnification.
Given sufficiently long light curves, the results will converge to the true
magnification offsets.  Even for Q2237, with its decade long OGLE light curves,
the data are not sampling long enough paths across the patterns
(see Figure \ref{fig:tracks}) to show convergence.  At present,
the distribution of differential mean magnification offsets are too broad
(Figure \ref{fig:zero}) to tightly constrain any systematic magnification offsets.
Fortunately, our results for the other physical parameters are little affected
by whether we allow these offsets to vary or constrain them with the extinction
estimates of \citet{Agol09}.

Finally, we show for the first time that microlensing variability in a 
lens gives the same results when analyzing different portions of its light curve.
The analysis of light curves LC1 and LC2, corresponding to the 1st and 2nd halves of the
11 year OGLE monitoring period, lead to statistically consistent distributions
for every parameter we consider.
This both confirms our ability to measure parameters and gives
us tighter constraints after combining the results.
It would be computationally challenging to analyze the full light curve
simultaneously because it becomes (exponentially?) harder to fit longer light curves.
However, such full analyses are likely needed for some quantities, particularly
the magnification offsets, to converge.

In the future we will likely include binaries, even though their effect is not likely
to influence the results other than interpreting the meaning of the mean stellar mass
(by up to $0.05M_\sun$, as discussed in \S\ref{sec:results}).
However, like the projection of our motion relative to the CMB,
the streaming velocities in Q2237 are small compared to the peculiar velocities,
and so are do little to break the degeneracy.
The effects of streaming velocities will be seen most strongly in true disk lenses
(none are known, except, potentially PMN~J2004$-$1349, \citet{Winn01}),
or in lenses such as Q~J0158$-$4325 (see \citet{Morgan08b} for a microlensing
analysis of this active system) lying close to the equator of the CMB dipole,
which will have the full $369\kms$ dipole motion \citep{Hinshaw09}.
These CMB equatorial lenses should also show significantly shorter
microlensing variability time scales.
Detecting this effect would be an independent confirmation of the kinematic
origin of the dipole.

Q2237 was a natural first candidate for a full analysis with moving stars
because of the excellent OGLE data,
short microlensing timescales, and negligible time delays between the images.
However, there is no problem extending our approach to analyzing microlensing data with
moving stars to any other microlensing analysis.  Even if the time delays are unknown,
cases with different trial delays could simply be tried sequentially \citep{Morgan08b}.
Moreover our method can easily be extended to multi-wavelength data sets to examine the
structure of the accretion disk varies with wavelength \citep{Poindexter08}.
The memory requirements would be too great
to fit each band simultaneously as in \citet{Poindexter08}, but we can use a modified
version of the method \citet{Dai09} applied to the joint
optical and X-ray analysis of RXJ1131$-$1231.
The models are first run on the band with the most epochs.
As good fitting trials are found, the starting points, velocities,
and $\chi^2$ matrix are saved.  Next, for each successive band, we
recompute the light curves corresponding to the epochs and source sizes
of the other wavelengths,
and the results of these new fits are used to continue the $\chi^2$ calculation.
Since the overall execution times are only modestly longer than using static stars,
there is no reason not to use this more physically correct approach.

\acknowledgments
SP thanks Jason Sawin for discussions on computational techniques.
We thank Tyoma Tuntsov for suggesting the inclusion of the parallax effect.
We thank Andrew Gould for providing the parallax code.
This work was supported in part by an allocation of computing time
from the Ohio Supercomputer Center.
This research was supported by NSF grant AST-0708082.

\end{document}